\documentclass[aps,twocolumn,amsmath, amssymb,prl,showpacs]{revtex4-1}
\usepackage{times}
\usepackage{graphicx}
\usepackage{psfrag}
\usepackage{ae}
\usepackage{amsmath,amssymb}

\begin{document}

\title{Enhanced flow of core-softened fluids through nanotubes}

\author{Jos\'e Rafael Bordin} \email{bordin@if.ufrgs.br} \affiliation{
  Instituto de F\'{\i}sica, Universidade Federal do Rio Grande do Sul,
  Caixa Postal 15051, CEP 91501-970, Porto Alegre, RS, Brazil}
\affiliation{Institut f\"ur Computerphysik, Universit\"at Stuttgart,
  Almandring 3, 70569 Stuttgart, Germany}

\author{Jos\'e S. Andrade Jr.} \email{soares@fisica.ufc.br}
\affiliation{Departamento de F\'{i}sica, Universidade Federal do
  Cear\'a, CEP 60451-970 Fortaleza, Cear\'a, Brazil}

\author{Alexandre Diehl} \email{diehl@ufpel.edu.br}
\affiliation{Departamento de F\'{\i}sica, Instituto de F\'{\i}sica e
  Matem\'atica, Universidade Federal de Pelotas, Caixa Postal 354, CEP
  96010-900, Pelotas, RS, Brazil}

\author{Marcia C. Barbosa} \email{marcia.barbosa@ufrgs.br}
\affiliation{Instituto de F\'{\i}sica, Universidade Federal do Rio
  Grande do Sul\\ Caixa Postal 15051, CEP 91501-970, Porto Alegre, RS,
  Brazil}

\begin{abstract}
  We investigate through non-equilibrium molecular dynamic simulations
  the flow of core-softened fluids inside nanotubes. Our results
  reveal a anomalous increase of the overall mass flux for nanotubes
  with sufficiently smaller radii. This is explained in terms of a
  transition from a single-file type of flow to the movement of an
  ordered-like fluid as the nanotube radius increases. The occurrence
  of a global minimum in the mass flux at this transition reflects the
  competition between the two characteristics length scales of the
  core-softened potential. Moreover, by increasing further the radius,
  another substantial change in the flow behavior, which becomes more
  evident at low temperatures, leads to a local minimum in the overall
  mass flux. Microscopically, this second transition results from the
  formation of a double-layer of flowing particles in the confined
  nanotube space. These special nano-fluidic features of core-softened
  particles closely resemble the enhanced flow behavior observed for
  liquid water inside carbon nanotubes.
\end{abstract}


\maketitle

The recent development of nanofabrication techniques have opened the
possibility of building nanoscale structures~\cite{Pr01} that can
effectively mimic biological channels~\cite{Ag01}. This new technology
has generated devices with application not only as controlled models
for biological systems, but also as laboratory gatekeepers that are
chemically selective and could therefore eventually function as
nanofilters~\cite{Ta12}.

At the nanoscale, however, fluid properties differ significantly from
what usual hydrodynamics would predict. For instance, this is the case
of recent experiments highlighting the exceptional properties of
carbon nanotubes~\cite{Ma05,Holt06,Wh08,Qin11}. Precisely, these
studies revealed that certain microelectromechanical fabrication
processes are capable to assemble a macroscopic collection of carbon
nanotubes with diameters in a range as small as $1.3$-$2$ nm.
Furthermore, it has been shown that the water flux in these special
membranes can be three to four orders of magnitude larger than the
value prediction from the continuum-based no-slip Hagen-Poiseuille
(HP) relation~\cite{Koles04,Wh08}, as also confirmed by computer 
simulations~\cite{Holt06}. 
This water flow increase is not uniform with the different nanotube radius.
As the nanotube radius is reduced, the flow decreases up to a certain
threshold, and for smaller radius it increases
again, as reported by computational~\cite{thomas08,thomas09}
and experimenal~\cite{Qin11} works. This non-monotonic behavior is
attributed to the transition from continuum to sub-continuum transport
as the nanotube shrinks~\cite{thomas09,Qin11} and can also be driven
by hydrophobicity~\cite{Le12}. 
However, despite the use of all atom models for water, 
the computational results are only qualitatively comparable to the experiments.
The approximations employed in the water models and friction effects might
be the origin of these numerical differences~\cite{Falk10}.
The different atomistic models for water give, therefore, a
qualitative description of the experimental enhancement flow. This
suggests that the major ingredient for the rapid flux is not related
to the details of the model, but to the effective interactions that
they do represent.

In the last two decades a number of effective model potentials have
been suggested for water~\cite{Ja98,Sc00a,Xu05,Oliveira06a,Barraz09,Silva10}.  
Despite their simplicity, these models exhibit in bulk the thermodynamic, 
dynamic and structural anomalies of water and also predict the existence 
of a second critical point hypothesized by Poole and collaborators for the 
ST2 water model~\cite{Po92}. In this way, we test if this effective models
for water-like fluids can qualitatively reproduce the anomalous enhanced flux in nanotubes
with small radius, observed in experiments for water~\cite{Qin11}. The presence of this anomalous
behavior will show that the enhanced flow will occur to liquids without any directionality, 
and can not depend on the formation and destruction of hydrogen bonds.

The fluid is modeled as point-like particles with effective
diameter $\sigma$ and mass $m$, interacting through the three
dimensional core-softened potential~\cite{Oliveira06a}

\begin{equation}
\frac{U(r_{ij})}{\epsilon} = 4\left[ \left(\frac{\sigma}{r_{ij}}\right)^{12} - 
\left(\frac{\sigma}{r_{ij}}\right)^6 \right] + 
u_0 {\rm{exp}}\left[-\frac{1}{c^2}\left(\frac{r-r_0}{\sigma}\right)^2\right]\;.
\label{AlanEq}
\end{equation}
where $r_{ij}=|\vec r_i - \vec r_j|$ is the distance between particles
$i$ and $j$. The first term on the right is the standard 12-6
Lennard-Jones (LJ) potential~\cite{AllenTild} and the second
corresponds to a Gaussian centered at $r_0/\sigma$, with depth
$u_0\epsilon$ and width $c\sigma$. For $u_0 = 5.0$, $c = 1.0$ and
$r_0/\sigma=0.7$, the potential (\ref{AlanEq}) displays two different
length scales, namely, one at $r_{ij}\approx \sigma$, where the force
has a local minimum, and another at $r_{ij} \approx 2 \sigma$, where
the fraction of imaginary modes shows a local maximum~\cite{Oliveira10}.  
It has been previously observed \cite{Oliveira06a} that the 
pressure-temperature phase diagram of this system at equilibrium 
exhibits features similar to the anomalies present in water~\cite{Kell67,Angell76}.

\begin{figure}[ht]
\begin{center}
\includegraphics[clip,width=8cm]{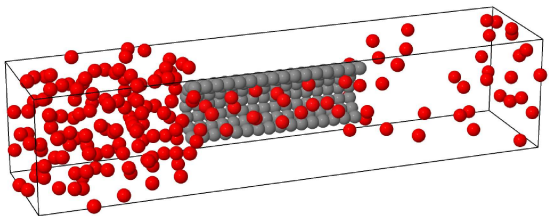}
\end{center}
\caption{Snapshot of the simulation box. The systems is composed of a high density reservoir, 
namely CV$_1$, and a low density reservoir, CV$_2$, connected by a 
nanotube. For better visualization, only half of the nanotube particles are shown. 
Flat and repulsive walls surround each entrance of the 
nanotube (not shown). The cylindrical nanotube in the center has radius
  $a$ and length $L_NT$.}
\label{fig1}
\end{figure}

Here we study the dynamic behavior of this fluid confined in a
nanotube connected to two reservoirs, namely, CV$_1$ on the left of
the nanotube and CV$_2$ at the right. An illustration of the
nanotube-reservoir setup is shown in Fig. \ref{fig1}. The simulation
box is a parallelepiped with dimensions $L_x\times L\times L$ in $x$-,
$y$ and $z$-directions, respectively. The box size in the
$x$-direction is defined as $L_x=4L$ for all simulations, and for the
$y-$ and $z$-directions is $L=10\sigma$ for all nanotubes with radius
$a\leq 4\sigma$, and $L=2a+2\sigma$ otherwise. The tube structure is
built as a wrapped sheet of LJ particles with diameter $\sigma_{\rm
  NT}=\sigma$. Nanotube and water-like particles interact through the
purely repulsive Weeks-Chandler-Andersen (WCA) potential
\cite{AllenTild}. 
The nanotube entrances are surrounded by flat walls, which also interact with
the fluid particles through the WCA potential.

The average temperature of the system is fixed by means of the
Nose-Hoover heat-bath scheme with a coupling parameter $Q=2$, and
periodic boundary conditions are applied in all directions. For
simplicity, we also assume that the nanotube atoms are motionless
(i.e., not time-integrated) during the entire simulation. A constant
time step of $\delta t=0.005$, in LJ time units~\cite{AllenTild} is
adopted.

\begin{figure}[ht]
\begin{center}
\includegraphics[width=8cm]{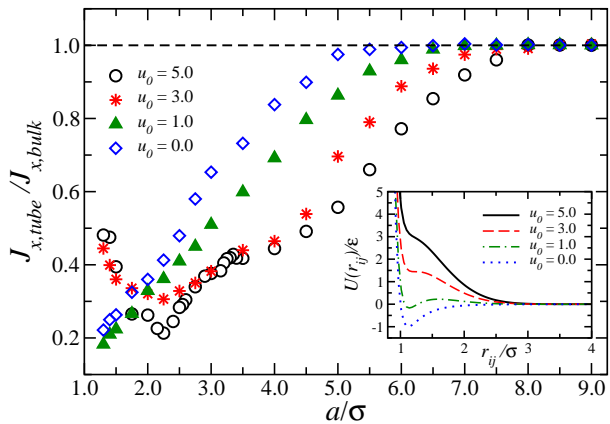}
\end{center}
\caption{Flux of particles through the nanotube, $J_{x,tube}$, in
  units of the non-confined flux, $J_{x,bluk}$, for different
  interaction potentials. The inset shows
  the fluid-fluid interaction, Eq.~(\ref{AlanEq}), as function of the separation distance
  $r_{ij} = |\vec r_i - \vec r_j|$ between two particles, and for different values of $u_0$.}
\label{fig2}
\end{figure}

A steady state flux through the nanotube is induced by fixing a higher
density in the reservoir CV$_1$, $\rho_1\sigma^3 = 0.2$, and a lower
density in CV$_2$, $\rho_2\sigma^3 = 0.01$. The initial configuration
of the particles in the system is generated using a standard GCMC
simulation in each reservoir, during $5\times10^5$ steps, with the
initial velocity for each particle obtained from a Maxwell-Boltzmann
distribution at a desired temperature. The density in the reservoirs
is maintained constant using the Dual Control Volume Grand Canonical
Molecular Dynamics (DCV-GCMD) method~\cite{DCVGCMD94,Bordin12a}.
Precisely, the desired densities are restored in CV$_1$ and CV$_2$ by
intercalating the MD steps with a number of GCMC steps inside the
corresponding control volumes, as depicted in Fig.~\ref{fig2}. In our
simulations, an initial $5\times10^5$ MD steps were used for
equilibration of the system, and 150 GCMC steps were performed for
every 500 MD steps during the DCV-GCMD process. Final results were
typically obtained after performing a total of $5\times10^7$ MD steps
for each simulation, and averaging over $10$ to $20$ independent runs to
evaluate relevant physical quantities.

The axial flux of particles through the tube, $J_{x,tube}$, is
computed by counting the number of particles that cross the channel
from left to right, $n_{ltr}$, and the particles flowing from right to
left, $n_{rtl}$ \cite{DCVGCMD94},
\begin{equation}
 J_{x,tube} = \frac{n_{\rm ltr}-n_{\rm rtl}}{A_{\rm NT}N_{\rm steps} \delta t},
\end{equation}
where $A_{\rm NT}=\pi a^2$, $N_{\rm steps}$ is the total number of
steps used in the simulation, and $\delta t$ is the MD time step. In
an entirely similar way, we evaluate the flux in the $x$-direction for
the non-confined case, $J_{x,bulk}$.

Here we assume that the fluid-fluid interaction, Eq.~(\ref{AlanEq}),
has a cutoff radius $r_{\rm cut}/\sigma = 3.5$ for all cases. The
nanotube radius is varied from $a/\sigma = 1.3$ to $a/\sigma=9.0$,
and, in all simulations, the tube length is fixed to
$L_{NT}/\sigma=15$.  In order to characterize the effect of the two
length scales in the potential (\ref{AlanEq}), simulations were
performed for four values of the parameter $u_0$, $u_0=5.0$, $3.0$,
$1.0$ and $0.0$, where this last case corresponds to the standard
$12-6$ LJ fluid.

Figure~\ref{fig2} illustrates the dependence of the mass flux with the
nanotube radius for different values of $u_0$. The inset shows how
this parameter affects the shape of the interaction potential. In all
models the temperature was fixed to $T^*=1.0$, where $T^* \equiv
k_BT/\epsilon$ and $k_B$ is the Boltzmann constant. For $u_0=5.0$ and
sufficiently large values of the nanotube radius, the flux reaches a
maximum which is approximately equal to its corresponding bulk value.
As expected, by decreasing the radius, the flux gradually decreases
from this saturation value, due to the obvious mass transport
limitations of the confined nanotube geometry. Interestingly, as the
radius is reduced even more, the flux reaches a minimum at
$a/\sigma\cong 2.25$ and starts to increase sharply, revealing
anomalously large values for a range of significantly reduced nanotube
radii. Anomalies in the transport behavior have also been previously observed with 
atomistic ~\cite{Zheng12}  and effective models~\cite{Bordin12b} of water. In these cases, the diffusion 
coefficient is evaluated in confined environments, but under equilibrium 
conditions, which leads to an apparent measure. This can be substantially different
from the effective mass transport $J_{x}$ quantified here.

As shown in the inset of Fig.~\ref{fig2}, the interaction potential
for $u_0=3.0$ displays a shoulder-like profile with a small barrier
between the two length scales. Our simulations for this case also
indicate the presence of a minimum followed by anomalously large flux
values for radii smaller than the critical radius, as can be seen in
Fig.~\ref{fig2}. Moreover, despite the decrease in the energy barrier
from $u_0=5.0$ to $u_0=3.0$, the critical radius is the same for both
potentials since they still possess identical characteristics lengths.

For $u_0=1.0$, the potential exhibits a quasi-LJ shape, as shown in
the inset of the Fig.~\ref{fig2}, with a small attractive well in the first
characteristic length and a small barrier between this and the second
characteristic length. As shown in Fig.~\ref{fig2}, the anomalous
behavior vanishes in this case. The competition between the two length
scales becomes irrelevant as it becomes energetically viable for a
particle to transit from positions corresponding to the first and
second characteristic lengths.
Previous works for this core-softened models show that, even in bulk, this
fluids only feature water anomalies when exist competition between the length
scales in the interaction potential~\cite{Oliveira09,Barraz09,Silva10}.
Finally, as expected for simple LJ liquids (i.e., $u_{0}=0$)~\cite{Striolo06},  we do not observe the 
anomalous increase in flux for small radius, as shown in Fig.~\ref{fig2}.
\begin{figure}[ht]
\begin{center}
\includegraphics[clip,width=8cm]{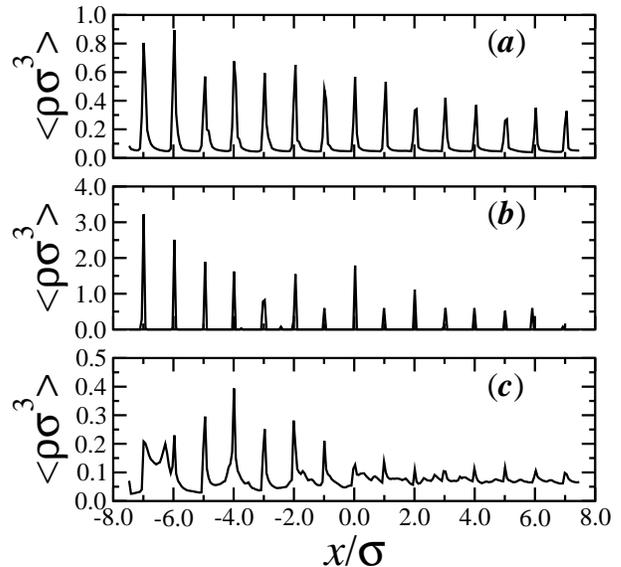}
\end{center}
\caption{Axial density profiles inside the nanotube for a fluid modeled 
with $u_{0}=5.0$ at temperature $T^{*}=1.0$.  An ordered
liquid behavior is observed for a nanotube with $a/\sigma=4.5$ (a), 
a solid-like behavior for $a/\sigma=2.0$ (b), and a disordered 
structure appears for $a/\sigma=1.25$ (c).}
\label{fig3}
\end{figure}

We can understand the dynamical behavior of the system through its
underlying structural properties. Here we consider typically the
simulation results obtained with $u_0 = 5.0$ at temperature $T^*=1.0$.
Fluids modeled with two-length scales potential tend to display an
ordered structure in bulk, which corresponds to the situation shown in
Fig.~\ref{fig3}a, where the confining nanotube has a large radius,
$a^* = 4.0$. The ordered behavior and the resulting high flux reflects
the quasi-discrete nature of the particle flow. The distance between
the two peaks of density is approximately equal to $\sigma$, the first
length scale of the core-softened potential. The fluid therefore
experiences a solid-liquid-like behavior due to changes in relative
positions of interacting particles from one density peak to another. 
The highly ordered density profile of the fluid observed in
Fig.~\ref{fig3}b at the critical nanotube radius is responsible for
the low particle flux. This solid-like state of minimum mobility
appears as a special condition due to a combined action of confinement
and the presence of two lengths scales in the potential. As shown in
Fig.~\ref{fig3}c, ordering disappears under extreme confinement to 
generate a single-file flow of enhanced flux, where the particles 
are obliged to remain at the first length scale. 
\begin{figure}[ht]
\begin{center}
\includegraphics[clip,width=8cm]{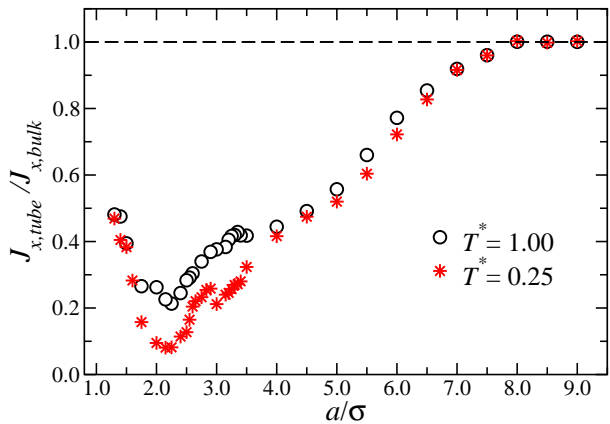}
\end{center}
\caption{Pparticle flux through the nanotube for the $u_0=5.0$
  case at different values of temperature. The anomalous enhanced flux in narrow nanotubes
  is observed for both temperatures, and a second anomalous behavior is clear 
  for low temperatures.}
  \label{fig4}
\end{figure}

\begin{figure}[ht]
\begin{center}
\includegraphics[clip,width=8cm]{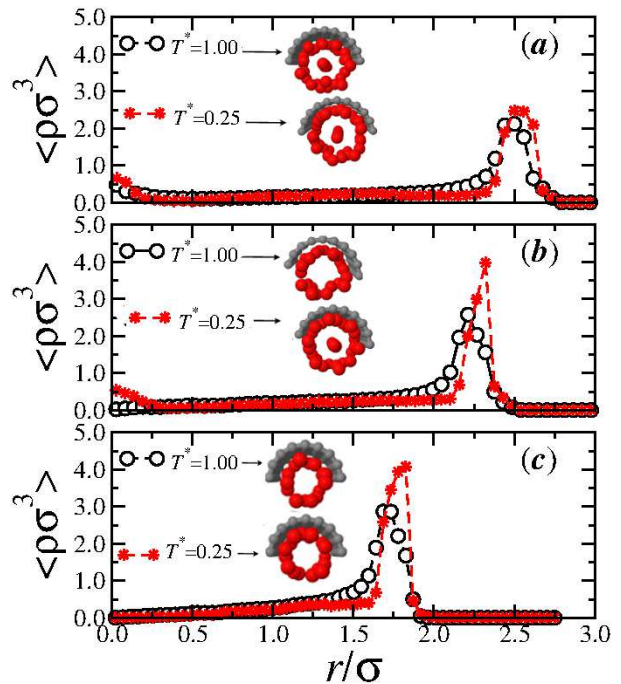}
\end{center}
\caption{Radial density profile inside the nanotube for $T^* = 1.0$
  (black circles) and $T^* = 0.25$ (red stars), and radius (a)
  $a/\sigma=3.5$, (b)$a/\sigma=3.25$ and (c) $a/\sigma=2.75$. These results indicate that
  the transition from a double layer structure to a single cylindrical layers
  is temperature dependent. The snapshots show the different structures.}
\label{fig5}
\end{figure}

Finally, we investigate the fluid flow behavior in the nanotube for
$u_0=5.0$ at different temperatures. A similar behavior is observed, 
with the minimum flux located at the same value of radius, as shown in
Fig~\ref{fig4}a for $T^*=1.0$ and $0.25$. 
Surprisingly, a second anomalous behavior is revealed at lower
temperatures. The system at $T^*=0.25$ shows a anomalous flux increase
in the region $3.0 \le a/\sigma \le 2.75$. Looking
in detail the curve for $T^*=1.0$,
the same behavior can be observed in the region $3.5 \le a/\sigma\le 3.12$.
This anomaly occurs due to changes in the fluid structure from a
double layer of flowing particles (i.e., a cylindrical layer near the
nanotube wall and a linear central layer) to a single cylindrical
layer structure. For nanotubes with radius $a/\sigma = 3.5$, the
system displays the same double layer structure for both temperatures,
as shown in Fig.~\ref{fig5}a. For $a/\sigma=3.25$, the single layer
structure appears at $T^*=1.0$, while the fluid at a lower temperature
$T^*=0.25$ still flows through a double layer structure, as depicted
in Fig.~\ref{fig5}b. This change in the fluid conformation causes the
observed flux anomaly, being a consequence of the competition between
the fluid-fluid and the fluid-wall interactions. 
The fluid particles tends to
form structures that increase the enthapic contribution to the
free energy, minimizing it. On the other hand, the confinement
imposes restrictions to ordering, increasing the entropic
contribution to the free energy~\cite{Bordin12b}.
As a consequence, the transition from the structure with two 
to one single cylindrical layer takes palce first in systems at higher temperature, 
or higher entropic contribution.
For sufficiently low values of the nanotube radius, the flowing system
is characterized by a single cylindrical layer, regardless of the
temperature, which is the case for $a/\sigma=3.25$, as shown in
Fig.~\ref{fig5}c.

In summary, here we studied the effect of confinement in the flow of
particles interacting through core-softened potentials. In the
presence of two competing length scales, our results show an anomalous
non-monotonic behavior in the form of a global minimum flux with the
nanotube radius, which is compatible with a transition from
single-file flow to the flow of an ordered-like fluid. At increased
values of the radius, a local minimum in the overall mass flux
appears, which is more evident at low temperatures, and reflects the
formation of a double-layer of flowing particles through the nanotube.
These anomalous flow properties of core-softened particles in confined
cylindrical geometries provide useful insight to the microscopical
description of the flow behavior of liquid water in carbon
nanotubes.
\section{Acknowledgments}
We acknowledge financial support from the Brazilian Agencies CNPq,
CAPES, FAPERGS, INCT-FCx and FUNCAP, the FUNCAP/CNPq Pronex grant, and the National
Institute of Science and Technology for Complex Systems in Brazil.
JRB thanks to CAPES for the financial support
for a collaborative period at Institute for Computational Physics,
University of Stuttgart - scholarship n$^{\rm o}$ 9155112.


\end{document}